\documentclass[a4paper,11pt]{article}
\usepackage[dvips]{graphicx}
\usepackage{amssymb}
\usepackage{amsmath}

\usepackage{cite}

\begin{document}

\title{Casimir Force of Fermions Coupled to Monopoles in Six Dimensional Spacetime}
\author{V. K. Oikonomou\thanks{
voiko@physics.auth.gr}\\
Max Planck Institute for Mathematics in the Sciences\\
Inselstrasse 22, 04103 Leipzig, Germany} \maketitle

\begin{abstract}
We calculate the Casimir force for a fermionic quantum field in a piston geometry with three parallel plates. The fermion satisfies bag boundary conditions on the plates and the spacetime is assumed to have compact extra dimensions. The calculation is performed in the cases where the extra space has toroidal and spherical topology. We are mainly interested in the case in which the fermion is coupled non-trivially to an extra dimensional defect, with a torus extra dimensional topological background. We found that in certain limits, the Casimir force corresponding to the defect-fermion system and to the sphere, has opposite sign, in reference to those corresponding to the toroidal extra dimensional spaces. 
\end{abstract}

\section*{Introduction}

The Casimir effect is an imprint of the quantum vacuum energy of a quantum field theory \cite{cas1,casimirbooks}. Since the theoretical prediction by Casimir in 1948 \cite{cas1}, many developments in the research area of the Casimir energy have been done, with the applications of the Casimir energy varying from the macro-scale physics \cite{fermioncasimirextradimensions,generalfermioncasimir,cosmologycasimir} to the micro-scale \cite{emig}, with the macro-scale physics having to do with cosmological predictions and the micro-scale application having to do with nano-scale and micro-scale mechanical devises. It is a well established fact that the Casimir energy is affected by the geometry and topology of spacetime and also drastically affected by the shape and the geometry of the apparatus. Since the Casimir energy is an important ingredient of a consistent quantum field theory, the calculations of the Casimir energy and force have been performed for scalar fields \cite{generalbosoncasimir}, fermion fields \cite{fermioncasimirextradimensions} for various spacetime topologies. Moreover, the topological structure of spacetime can be revealed or be severely constrained by sophisticated Casimir force experiments \cite{casimirexper}.   

\noindent Particularly, since the Casimir energy experiments are much less expensive in comparison to particle collider experiments, they can be used as a mean to reveal the non-trivial low energy spacetime structure, with the non-triviality being caused by extra dimensions for example. In reference to extra dimensions, a lot of questions have to be answered, for instance, what is the compactification mechanism, how extra dimensions are stabilized, what is the compactification scale, how many extra dimensions exist, what is their topology and geometry etc. Since string theory predictions involve many extra dimensional compactifications for matter fields (that is fermions fields), with some of these compactifications involving non-trivial fluxes or defects, it is interesting to find the impact of such compactification mechanisms in the low energy physical phenomena. This actually serves as a motivation to study theoretically the Casimir energy of fermion fields in the presence of such extra-dimensional non-trivial configurations, and calculate the Casimir force on a macroscopic system, such as two or more parallel plates.  

\noindent One refined technique for calculating the Casimir force is that of the Casimir piston, which has received considerable attention in the last years. For an important stream of papers on Casimir pistons see \cite{pistons}. This is owing to the attractive quantitative features that the
piston setup has. The Casimir piston is materialized by three parallel
plates, with a quantum field considered to exist between these plates. Thus, the calculation of Casimir force on a
piston in the presence of compact extra dimensions is an important task in order to see the effect of the extra dimensions in such configurations. The piston geometry and in general, multiple parallel plates periodic geometry is frequently realized in nano and micro-devices. Due to the attractive attributes that the piston setup offers, we shall study the fermionic field Casimir force for the piston setup, in the presence of extra dimensions. Particularly, we shall calculate the Casimir force for massless fermionic fields coupled to a defect in the extra dimensional space. This is a low energy analogue of the superstring inspired flux compactifications. Moreover, the fermions are assumed to satisfy bag boundary conditions on the piston plates, which is the only form of boundary conditions appropriate for fermions, since the Dirac equation cannot be solved in corners (see \cite{oikonotrac} and references therein). In order to have a clear picture of the behavior of the Casimir force, we shall also calculate the Casimir force for various other extra dimensional spaces and compare the results to the defect-fermion case. As we will demonstrate, in certain limiting cases, the Casimir force corresponding to the defect-fermion and sphere cases behaves completely differently, in reference to the toroidal extra dimensional spaces case.

\noindent This paper is organized as follows: In section 1, we present the piston setup that will be used in the forthcoming sections and provide some general expressions of the Casimir energy and force, in the piston setup. In section 2 we calculate the Casimir force for the defect-fermion system with a torus extra dimensional space and the defect existing in the extra dimensions. The calculation is performed in various limiting cases. The same analysis is done in section 3, in the case  the extra dimensional space is a sphere, torus and a circle, with various boundary conditions for the fermions in the extra space. A discussion along with the conclusions follow in section 4.

\section{Casimir Piston Geometry and the Fermionic Casimir Energy in the Presence of Extra Dimensions}

\begin{figure}[t]
\begin{center}
\includegraphics[scale=.8]{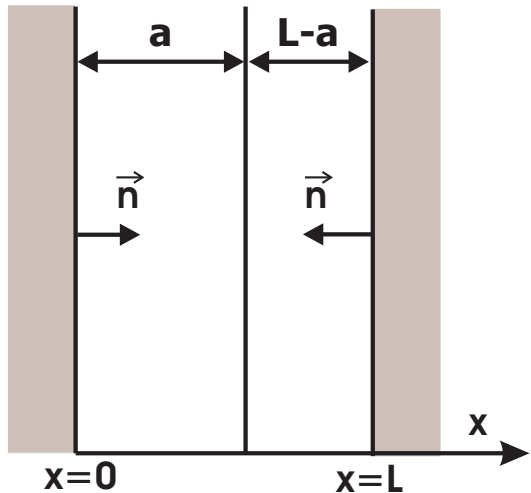}
\end{center}
\caption{The Casimir piston in the $x$ direction}
\label{olaskata3dfgdfg}
\end{figure}
Assume that the total spacetime time has the form $\mathcal{M}^4\times \mathcal{X}$, with $\mathcal{M}^4$ denoting the four dimensional Minkowski spacetime and $\mathcal{X}$ the extra dimensional space. The piston setup is like the one in Fig. (\ref{olaskata3dfgdfg}), where $x$ denotes one of the three spatial dimensions of the four dimensional Minkowski spacetime. The piston has two chambers with lengths $L-a$ and $a$, and the fermion field is assumed to be confined in each chamber, and also that it satisfies bag boundary conditions on the plates at $x=0$, $x=a$ and at $x=L$. Let us briefly recall the essentials of bag boundary conditions. Consider a Dirac fermion field in 3 space dimensions with three parallel plates in one of the three spatial dimensions, at $x=0$, $x=a$ and at $x=L$. We shall also assume that fermions are not allowed to exist outside the parallel plate system. This is the physical implication of the MIT bag
boundary conditions which are expressed by the equation,
\begin{equation}\label{mitbag}
in^{\mu}\gamma_{\mu}\psi=\psi
\end{equation}
or in Lorentz covariant form,
\begin{equation}\label{mitbagcov}
n^{\mu}\bar{\psi}\gamma_{\mu}\psi=0
\end{equation}
where $n^{\mu}=(0,\vec{n})$ and $\vec{n}$ is the vector normal to the
surface of the plates directed to the interior of the slab
configuration. The above two equations indicate that there is no
fermion current flowing outwards from the parallel plates. The eigenvalues of the massless Dirac equation
obeying bag boundary conditions between two plates at $x=0$ and $x=L$ are \cite{oikonotrac},
\begin{equation}\label{dhyf}
\omega_n=\sqrt{k^2+\frac{\pi^2(n+\frac{1}{2})^2}{L^2}}
\end{equation}
where ''$k$'' refer to the transverse components of the
momentum. We shall provide a general expression of the Casimir energy using the zeta-regularization method \cite{generalcasimirzetaregularization}. In addition, we adopt the dimensional regularization technique, by assuming $D$ Euclidean spatial dimensions. The Dirac fermion Casimir energy for a massless fermion reads,
\begin{align}\label{vlad}
&\mathcal{E}_c(s,a)= -2\sum_{n=0}^{\infty}\sum_{E_{KK}}\int
\frac{\mathrm{d}^{D-1}k}{(2\pi)^{D-1}}\Big{[}k^2+\frac{(n+\frac{1}{2})^2\pi^2}{a^2}+E_{KK}\Big{]}^{-s}
\end{align}
where the factor two indicates the spin and particle multiplicity, and $E_{KK}$ is the square of the eigenvalues of the Dirac operator for the compact extra dimensional space. The summation over the $E_{KK}$ is done over the quantum numbers characterizing the extra dimensions. Note that in the end we must substitute $s=-\frac{1}{2}$ and
$D=3$. Upon integrating over the continuous dimensions using the formula,
\begin{equation}\label{feynman}
\int
\mathrm{d}k^{D-1}\frac{1}{(k^2+A)^s}=\pi^{\frac{D-1}{2}}\frac{\Gamma(s-\frac{D-1}{2})}{\Gamma(s)}\frac{1}{A^{s-\frac{D-1}{2}}}
\end{equation}
the Casimir energy (\ref{vlad}) reads,
\begin{align}\label{pordoulis}
&E_c(s,a)=-\frac{1}{2\pi^{\frac{D-1}{2}}}\frac{\Gamma(s-\frac{D-1}{2})}{\Gamma(s)} \sum_{E_{KK}} \sum_{n=0}^{\infty}\Big{[}\frac{(n+\frac{1}{2})^2\pi^2}{a^2}+E_{KK}\Big{]}^{-(s-\frac{D-1}{2})}.
\end{align}
The above relation can be written in terms of the inhomogeneous
Epstein zeta function
\cite{casimirbooks},
\begin{equation}\label{cativoepsteinzeta}
\mathcal{Z}_d^{v^2}\Big{(}s;w_1,...,w_d,c_1,...,c_d\Big{)}=\sum_{n_1...n_N=-\infty}^{\infty}[w_1(n_1+c_1)^2+...+w_d(n_d+c_d)^2+v^2]^{-s}.
\end{equation}
as follows,
\begin{align}\label{pordoulis13}
&\mathcal{E}_c(s,a)=-\frac{1}{2\pi^{\frac{D-1}{2}}}\frac{\Gamma(s-\frac{D-1}{2})}{\Gamma(s)} \sum_{E_{KK}} \mathcal{Z}_1^{E_{KK}}\Big{(}s-\frac{D-1}{2};\frac{\pi^2}{a^2};\frac{1}{2}\Big{)}
\end{align}
Using the formula,
\begin{align}\label{gone}
&\sum_{n=0}^{\infty
'}\Big{(}a(n+\frac{1}{2})^2+q\Big{)}^{-s}
=-\sqrt{\frac{\pi}{a}}\frac{\Gamma(s-\frac{1}{2})}{2\Gamma(s)}{\,}\frac{q^{-s+\frac{1}{2}}}{2} 
+\frac{2\pi^sq^{-\frac{s}{2}+\frac{1}{4}}a^{-\frac{s}{2}-\frac{1}{4}}}{\Gamma(s)}
\\& \notag \times \sum_{n=1}^{\infty}(-1)^nn^{s-\frac{1}{2}}K_{s-\frac{1}{2}}\Big{(}2\pi
n\sqrt{\frac{q}{a}}\Big{)}
\end{align} 
and also,
\begin{equation}
 \sum_{n=1}^{\infty}(-1)^qf(r)=2\sum_{n=1}^{\infty}f(2r)-\sum_{n=1}^{\infty}f(r)
\end{equation}
the Casimir energy can be cast as:
\begin{align}\label{pex1}
&\mathcal{E}_c(s,a)=-\frac{1}{(2\pi)^{D-1}\Gamma(s)} \sum_{E_{KK}}\Big{[} -\frac{a}{2\sqrt{\pi}}\Gamma(s-\frac{D-2}{2})E_{KK}^{-(s-\frac{D-1}{2})+\frac{1}{2}} 
\\& \notag+\frac{2\pi^{s-\frac{D-1}{2}}E_{KK}^{-\frac{1}{2}(s-\frac{1}{2})+\frac{1}{4}}\pi^{-\frac{1}{2}(s-\frac{1}{2})-\frac{1}{4}}}{a^{-\frac{1}{2}(s-\frac{1}{2})-\frac{1}{4}}}\\ \notag & \times (2\sqrt{E_{KK}}a)^{-s+\frac{D-2}{2}}\Big{(}2\sum_{n=1}^{\infty}(4\sqrt{E_{KK}}na)^{s-\frac{D-2}{2}}K_{s-\frac{D-2}{2}}\Big{(}4
n\sqrt{E_{KK}}a\Big{)}
\\ \notag &
-\sum_{n=1}^{\infty}(2\sqrt{E_{KK}}na)^{s-\frac{D-2}{2}}K_{s-\frac{D-2}{2}}\Big{(}2
n\sqrt{E_{KK}}a\Big{)}\Big{]}
\end{align}
Notice the first term on the right hand side. It is singular, since it contains gamma function which is singular for $D=3$. The contribution from the chamber $L-a$, can be found from the above by simply making the replacement:
\begin{equation}\label{rep}
a\rightarrow L-a
\end{equation}
Consequently, the total Casimir energy of the piston, is the sum of the two contributions,
\begin{equation}\label{toto1}
\mathcal{E}_{piston}=\mathcal{E}(s,a)+\mathcal{E}(s,L-a)
\end{equation}
Therefore, the Casimir force is equal to:
\begin{equation}\label{fp}
F_c=-\frac{\partial \mathcal{E}_{piston}}{\partial a}
\end{equation}
Observe that the final expression of the Casimir force is regular, since the singular terms canceled each other when the contributions from the two chambers were added. This is the greatest attribute of the Casimir piston geometry. In the following sections we shall study analytically the various limiting cases of the Casimir force for the defect-fermion system, and compare it to the Casimir force corresponding to other extra dimensional compact spaces.

\subsection{Small Argument Expansion of the Casimir Force for the Piston Geometry}

Before we close this section, we shall perform a small argument approximation for the Casimir force and provide a general formula for the force. We shall use this general formula quite frequently in the sections that follow. In order to find the small argument expansion of the Casimir force, we write the Casimir energy in the following form:
\begin{align}\label{pordoulissme}
&E_c(s,a)=-\frac{1}{2\pi^{\frac{D-1}{2}}}\frac{\Gamma(s-\frac{D-1}{2})}{\Gamma(s)}\Big{(}\frac{a^2}{\pi^2}\Big{)}^{s-\frac{D-1}{2}} \sum_{E_{KK}} \sum_{n=0}^{\infty}\Big{[}(n+\frac{1}{2})^2+\frac{E_{KK}a^2}{\pi^2}\Big{]}^{-(s-\frac{D-1}{2})}.
\end{align}
We shall use the small-q expansion, which can be found in the book of Elizalde \cite{casimirbooks} (page 83 equation 4.34), 
\begin{equation}\label{smallq}
\sum_{n=0}^{\infty}\Big{[}(n+c)^2+q\Big{]}^{-s}=\sum_{m=0}^{\infty}\frac{(-1)^m\Gamma (m+s)}{\Gamma (s)m!}q^m\zeta_H(2s+2m,c)
\end{equation}
where $\zeta_H$ is the Hurwitz zeta function,
\begin{equation}\label{hurwitzzeta}
\zeta_H (s,b)=\sum_{N=0}^{\infty}\frac{1}{(N+b)^s}
\end{equation}
Therefore, the Casimir energy for the $a$-chamber is equal to:
\begin{align}\label{energysmallqrvfb}
& E_c(s,a)=-\frac{1}{2\pi^{\frac{D-1}{2}}}\frac{1}{\Gamma(s)}\Big{(}\frac{a^2}{\pi^2}\Big{)}^{s-\frac{D-1}{2}}\times 
\\ \notag & \sum_{E_{KK}}\sum_{m=0}^{\infty}\frac{(-1)^m\Gamma (m+s-\frac{D-1}{2})}{m!}\frac{E_{KK}^m}{\pi^{2m}}a^{2m}\zeta_H\Big{(}2(s-\frac{D-1}{2})+2m,\frac{1}{2}\Big{)}
\end{align}
Differentiating with respect to ''a'' and adding the contributions of the two chambers, we get for $D=3$ and $s=-1/2$:
\begin{align}\label{forcesmallq}
& F_c(s,a)=-\frac{5}{2\Gamma(-1/2)}\times
\\ \notag & \sum_{m=0}^{\infty}\frac{(-1)^m\Gamma (m-\frac{3}{2})}{(m-1)!}\sum_{E_{KK}}E_{KK}^m\zeta_H\Big{(}2m-3,\frac{1}{2}\Big{)}\Big{(}(L-a)^{2m-4}-a^{2m-4}\Big{)}
\end{align}
For $m=2$ the Hurwitz zeta function becomes singular, but the expression (\ref{forcesmallq}) is regular because the singularities from the two chambers cancel. This is exactly what was expected, owing to the piston geometry. Utilizing equation (\ref{forcesmallq}), we shall find in the following sections the small argument expansion for various extra dimensional spaces. The small argument expansion for our case can be satisfied when the following constraints hold true:
\begin{align}\label{valuesconstraints}
& a\ll 1 \\ \notag &
\frac{(L-a)^2}{R^2}\ll 1 \\ \notag &
a\ll R
\end{align}
For the purposes of this article and according to the above constraints, suffices to take $L\sim 10^{-7}$m and $R\sim 40\mu m$, with the latter choice being consistent with the Newton's law experiments.

\section{Six Dimensional Casimir Effect for Orbifold Torus Coupled to a Defect}

The defect-fermion system that we shall use has been studied in \cite{segre}. It worths outlining the general features of this system, in order to render the article self-contained. Following \cite{segre}, the defect in the extra dimensional two torus, is a  monopole, and is constructed in the following way. Consider a sphere $S^2$ described by the polar coordinates $(y_1, y_2)$, with $0 \leq y_2< 2\pi$ and $0\leq y_1 \leq \pi$. Two patches are needed on this manifold, the upper and the lower hemisphere with intersection at the equator $y_1 =\frac{\pi}{2}$,
\begin{align}\label{hemispheres}
& H_+: 0 \leq y_1 \leq \frac{\pi}{2},{\,}{\,}{\,} 0 \leq y_2< 2\pi 
\\ \notag & H_+: \frac{\pi}{2} \leq y_1 \leq \pi,{\,}{\,}{\,} 0 \leq y_2 < 2\pi
\end{align}
The matter field phases on the two coordinate patches, must be related as:
\begin{equation}\label{phasepatche}
\phi_+(y_1, y_2)=e^{in_my_2}\phi_{-}(y_1, y_2 )
\end{equation}
The number $n_m$ is an integer, since the phases must be single valued on the equator, and this number is actually the winding number of the large gauge transformation. The gauge potentials are related by the following gauge transformation:
\begin{equation}\label{lgt}
A_+=A_{-}+n_m\mathrm{d}y_2, 
\end{equation}
with:
\begin{equation}\label{diracpotemt}
A_{\pm}=\frac{n_m}{2}(\cos y_1\mp 1)\mathrm{d}y_2
\end{equation}
The torus $T^2$, can be constructed from $S^2$ by restricting the coordinate $y_1$ to the range $a\leq y_1 \leq  \pi -a$ and then equating $y_1 =a$ with $y_1= \pi-a$. We assume that the circumferences of the two circles that constitute the torus are equal, and $0\leq y_{1,2}\leq R$. The relations between the potentials and matter fields on the patches $H_{\pm}$, are given by:
\begin{align}\label{relations}
&A_+(y_1,y_2)=A_{-}(y_1,y_2)+\frac{2\pi n_m}{R}\mathrm{d}y_2
\\ \notag & \phi_+(y_1, y_2)=e^{\frac{2\pi i n_my_2}{R}}\phi_{-}(y_1, y_2 )
\end{align}
The following boundary conditions are imposed:
\begin{equation}\label{bcimpo}
A'_+(0,y_2)=A'_{-}(R,y_2),{\,}{\,}{\,}\phi'_{+}(0,R)=\phi'_{-}(R,y_2)
\end{equation}
When expressed as coordinate conditions of the functions on the ``$+$'' patch, these are written as follows,
\begin{align}\label{pluspatch}
&A'_+(R,y_2)=A'_{+}(0,y_2)+\frac{2\pi n}{R}\mathrm{d}y_2, 
\\ \notag &\phi_+'(R, y_2)=e^{\frac{2\pi i n_my_2}{R}}\phi_{+}'(0, y_2 )
\end{align}
Therefore, an appropriate choice of the potential, which satisfies the above conditions and also constitutes a monopole on the space $T^2$, with winding number $n_m$, is:
\begin{align}\label{convchoiceofpotential}
A'_{+}(y_1,y_2)=\frac{2\pi n_m}{R^2}y_1\mathrm{d}y_2
\end{align}
The field strength reads,
\begin{equation}\label{noname}
F=\mathrm{d}A'_{\pm}=\frac{2\pi n_m}{R^2}\mathrm{d}y_1\wedge \mathrm{d}y_2
\end{equation}
The Dirac fermion Lagrangian, coupled to the monopole field is equal to:
\begin{equation}\label{lagrang}
\mathcal{L}=-\frac{1}{4}F_{M ,N }F^{M ,N }+\frac{1}{2}(\bar{\Psi)}\Psi\Gamma^M D_M\Psi-D_M\bar{\Psi}\Gamma^M\Psi,
\end{equation}
with $D_M=\partial_M-ieA_M$. The total spacetime manifold $\mathcal{M}^4\times T^2$ is described by the coordinates $(x^{\mu},y_1,y_2)$ and the metric is assumed to have the signature $(+,---;--)$. Obviously the total manifold is flat. The Dirac gamma matrices are decomposed as a tensor product:
\begin{align}\label{gammadirac}
& \Gamma^{\mu}\otimes I_2
\\ \notag & \Gamma^5=i\gamma^5\otimes \tau^1
\\ \notag & \Gamma^6=i\gamma^5\otimes \tau^2
\end{align}
with $\gamma^{\mu}$ are the four dimensional matrices, and $\tau^{1,2}$ are the Pauli matrices. The gauge background is of the form:
\begin{equation}\label{bagauge}
e A_M(x^{\mu},y_1,y_2)=\Big{(}0,0,0,0;0,\frac{2\pi n_m}{R^2}\Big{)}
\end{equation}
As already mentioned, the electromagnetic field strength has only one non-zero component, namely:
\begin{equation}\label{nonzeromod}
eF_{56}\mathrm{d}y_1\wedge \mathrm{d}y_2=\frac{2\pi n_m}{R^2}\mathrm{d}y_1\wedge \mathrm{d}y_2
\end{equation}
which is of the form (\ref{noname}).

\begin{figure}[h]
\begin{minipage}{15.5pc}
\includegraphics[width=17pc]{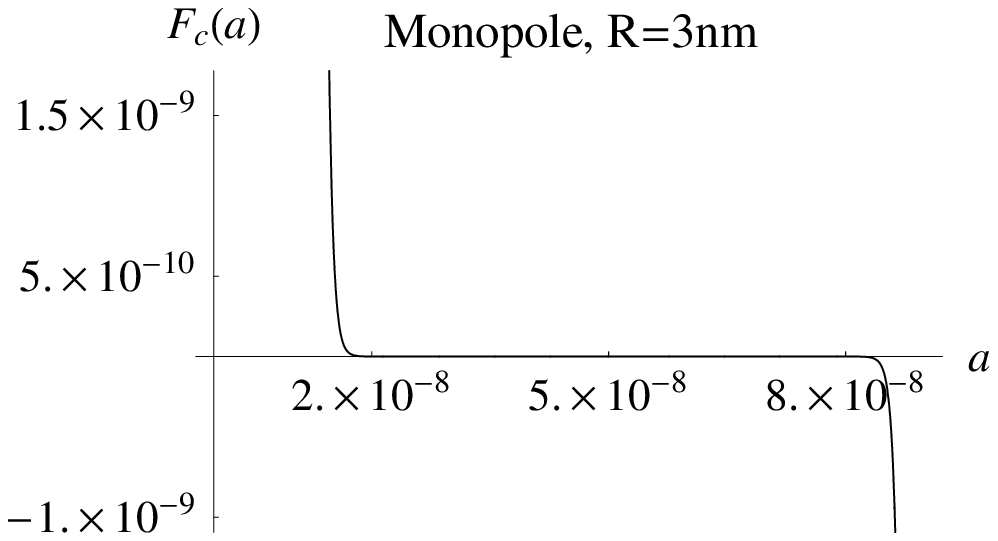}
\end{minipage}
\begin{minipage}{28pc}\hspace{2pc}%
\includegraphics[width=17pc]{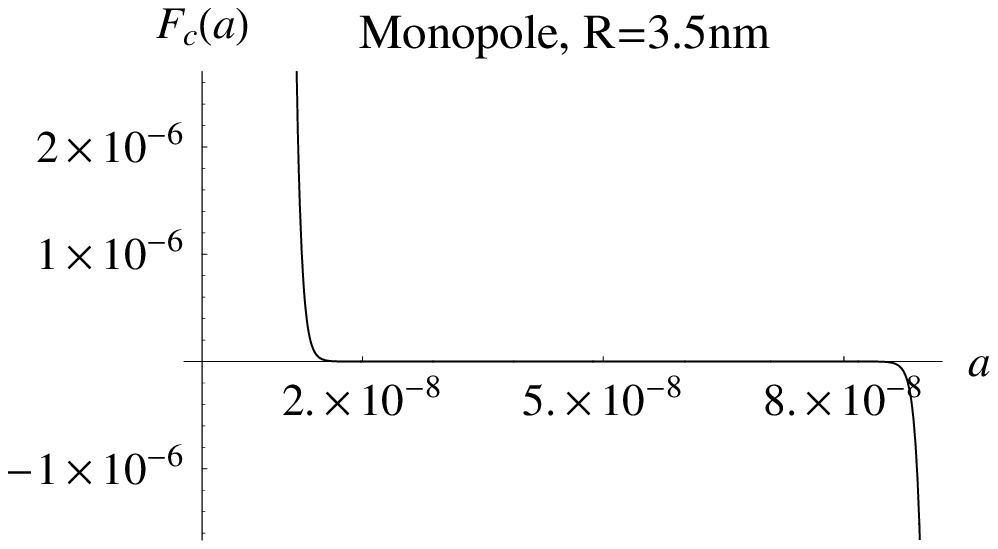}
\end{minipage}
\begin{minipage}[b]{14pc}
\caption{\label{olaskata1}The Casimir force on the central plate of a piston, as a function of $a$ for $R=3$nm and $R=3.5$nm. }
\end{minipage}
\end{figure}

\noindent In order to calculate the Casimir energy, we must find the eigenmodes of the Dirac operator in this torus-monopole background. These can be obtained by solving the following equation,
\begin{equation}\label{solvediraceqn}
i\gamma^{k}\partial_{k}\Psi_j(x^{\nu},y_1,y_2)=\lambda_j\Psi_j
\end{equation}
with $\lambda_j$, the eigenvalues of the Dirac operator and $k=0,1,2,...,6$. This task is simplified if we find the eigenvalues of the following operator:
\begin{equation}\label{solvediraceqn1}
\Big{(}i\gamma^{k}\partial_{k}\Big{)}^2\Psi_j(x^{\nu},y_1,y_2)=E_N\Psi_j
\end{equation}
with $\Big{(}i\gamma^{k}\partial_{k}\Big{)}^2$, being equal to:
\begin{equation}\label{newop}
\Big{(}i\gamma^{k}\partial_{k}\Big{)}^2=\Big{[}-\partial_{\mu}\partial^{\mu}+\frac{\partial}{\partial y_1^2}+\Big{(}\frac{\partial }{\partial y_2}-\frac{2i\pi n_m}{R^2}y_1\Big{)}^2+\frac{2\pi n_m}{R^2}\tau^3\Big{]}
\end{equation}
Solving the above, the square of the eigenvalues of the Dirac operator in such a background are obtained \cite{segre}:
\begin{equation}\label{encaseorbi}
E_{KK}=\frac{4\pi}{R^2}n_m\Big{(}N+\frac{1}{2}\Big{)}
\end{equation}
with $N=0,1,...$. These are the Kaluza-Klein modes of the Dirac operator that enter the Casimir energy calculation. Substituting these in relation (\ref{pex1}), the  Casimir energy for the chamber with length $a$ reads,
\begin{align}\label{pex111}
&\mathcal{E}_c(s,a)=-\frac{2\pi^{\frac{D-1}{2}}}{(2\pi)^{D-1}\Gamma(s)} \sum_{N=0}^{\infty}\Big{[} -\frac{a}{2\sqrt{\pi}}\Gamma(s-\frac{D-2}{2})\Big{(}\frac{4\pi}{R^2}n_m\Big{(}N+\frac{1}{2}\Big{)}\Big{)}^{-(s-\frac{D-1}{2})+\frac{1}{2}} \\ \notag &+\frac{2\pi^{s-\frac{D-1}{2}}\Big{(}\frac{4\pi}{R^2}n_m\Big{(}N+\frac{1}{2}\Big{)}\Big{)}^{-\frac{1}{2}(s-\frac{1}{2})+\frac{1}{4}}\pi^{-\frac{1}{2}(s-\frac{1}{2})-\frac{1}{4}}}{a^{-\frac{1}{2}(s-\frac{1}{2})-\frac{1}{4}}}\\ \notag & \times \Big{(}2\sqrt{\Big{(}\frac{4\pi}{R^2}n_m\Big{(}N+\frac{1}{2}\Big{)}\Big{)}}a\Big{)}^{-s+\frac{D-2}{2}}\Big{(}2\sum_{n=1}^{\infty}\Big{(}4\sqrt{\frac{4\pi}{R^2}n_m\Big{(}N+\frac{1}{2}\Big{)}}na\Big{)}^{s-\frac{D-2}{2}}
\\ \notag & \times K_{s-\frac{D-2}{2}}\Big{(}4
n\sqrt{\frac{4\pi}{R^2}n_m\Big{(}N+\frac{1}{2}\Big{)}}a\Big{)}
\\ \notag &
-\sum_{n=1}^{\infty}\Big{(}2\sqrt{\frac{4\pi}{R^2}n_m\Big{(}N+\frac{1}{2}\Big{)}}na\Big{)}^{s-\frac{D-2}{2}}K_{s-\frac{D-2}{2}}\Big{(}2
n\sqrt{\frac{4\pi}{R^2}n_m\Big{(}N+\frac{1}{2}\Big{)}}a\Big{)}\Big{]}
\end{align}
Analogously, we can obtain the contribution for the $L-a$ chamber, but we omit it for brevity. It is much more convenient to work in limiting cases of the theory. In the approximation where the argument of the Bessel function that appear in the fourth and fifth line of relation (\ref{pex1}) is large (we shall refer to this as ``large argument approximation'' in the following),  the Casimir force is equal to:
\begin{align}\label{Casflargemonopolegeneral1}
&\mathcal{F}_c\simeq \sum_{N=0}^{\infty}\frac{2\pi^{\frac{s-D+1}{2}}\Big{(}\frac{4\pi}{R^2}n_m\Big{(}N+\frac{1}{2}\Big{)}\Big{)}^{\frac{-4s+2D-2+1}{8}}a^{\frac{-4s-2D-3}{8}}}{\sqrt{n}}n^{s-\frac{D-2}{2}}
 \\  \notag & \times \Big{[}\Big{(}\frac{-4s-2D-3}{8}\Big{)}\frac{1}{a}\Big{(}2^{s-\frac{D-2}{2}+\frac{1}{2}}\exp\Big{(}-
4n\sqrt{\frac{4\pi}{R^2}n_m\Big{(}N+\frac{1}{2}\Big{)}}a\Big{)}
\\  \notag & -\exp\Big{(}-
2n\sqrt{\frac{4\pi}{R^2}n_m\Big{(}N+\frac{1}{2}\Big{)}}a\Big{)}
\\  \notag & -\frac{1}{L-a}\Big{(}\frac{-4s-2D-3}{8}\Big{)}-\Big{(}2^{s-\frac{D-2}{2}+\frac{1}{2}}\exp\Big{(}-
4n\sqrt{\frac{4\pi}{R^2}n_m\Big{(}N+\frac{1}{2}\Big{)}}(L-a)\Big{)}
\\  \notag &-\exp\Big{(}-
2n\sqrt{\frac{4\pi}{R^2}n_m\Big{(}N+\frac{1}{2}\Big{)}}(L-a)\Big{)}\Big{)}
 \\& \notag-\Big{(}2^{s-\frac{D-2}{2}
+\frac{1}{2}}4n\sqrt{\frac{4\pi}{R^2}n_m\Big{(}N+\frac{1}{2}\Big{)}}\exp\Big{(}-
4n\sqrt{\frac{4\pi}{R^2}n_m\Big{(}N+\frac{1}{2}\Big{)}}a\Big{)}
\\  \notag &-2n\sqrt{\frac{4\pi}{R^2}n_m\Big{(}N+\frac{1}{2}\Big{)}}\exp\Big{(}-
2\sqrt{\frac{4\pi}{R^2}n_m\Big{(}N+\frac{1}{2}\Big{)}}\Big{)}
\\  \notag & -\Big{(}2^{s-\frac{D-2}{2}+\frac{1}{2}}4n\sqrt{\frac{4\pi}{R^2}n_m\Big{(}N+\frac{1}{2}\Big{)}}\exp\Big{(}-
4n\sqrt{\frac{4\pi}{R^2}n_m\Big{(}N+\frac{1}{2}\Big{)}}(L-a)\Big{)}
\\  \notag & -2n\sqrt{\frac{4\pi}{R^2}n_m\Big{(}N+\frac{1}{2}\Big{)}}\exp\Big{(}-
2n\sqrt{\frac{4\pi}{R^2}n_m\Big{(}N+\frac{1}{2}\Big{)}}(L-a)\Big{)}\Big{)}\Big{]}
\end{align}
where we used the following approximation for the Bessel function, which holds for large values of the argument ``$z$'':
\begin{equation}\label{Besselapprox}
K_{s}(z)\simeq \sqrt{\frac{\pi}{2z}}e^{-z} 
\end{equation}
Due to the exponential dependence of the Casimir force, it is reasonable to keep only the lowest terms in the summations, that is $n_m=1$ and $N=0$, $n=1$. In order to have an idea how this force behaves as a function of the compactification radius and the lengths $L$, $a$, we have to give some realistic values to these, always compatible to the approximation (\ref{Besselapprox}). Let us recall some facts from conventional Casimir energy experiments for parallel plates geometry. In the plate geometry, what is measured is the force per surface, which is of the form \cite{casimirbooks}:
\begin{equation}\label{forcepersurface}
\mathcal{F}_c=-\frac{\pi^2\hslash c}{240a^4}S
\end{equation}
with $a$ the distance between the two plates and $c$ the speed of light. Moreover, it is assumed that $S\gg a^2$. For a realistic experiment, $a$ is of the order of $1\mu m$, and the surface $S\sim 1\mathrm{cm}^2$. In such a case, the Casimir force is of the order of $10^{-7}$ N. For the case at hand we shall assume that $L\sim 0.1\mu $m, and search when the large argument approximation of the Bessel function is valid. As it can be easily checked, the approximation is valid when the compactification radius of the extra dimensional space is $R \leq 5 \times 10^{-9}$nm. These values are certainly compatible to the constraints posed by Newton law experiments on the compactification radius, in which experiments the compactification radius has to be $R<0.05$mm \cite{oikonomounewton,casimirbooks}. Supposing that $S=0.01\mathrm{cm}^2$ and also that $a$ always takes values compatible to the approximation (\ref{Besselapprox}), we plot in Fig. (\ref{olaskata1}), the dependence of the Casimir force on the central plate, as a function of $a$, for the cases that the compactification radius is $R=3$nm and $R=3.5$nm. Notice that a slight change of the compactification radius $R$ has a severe impact on the scale of the Casimir force, rendering it very small or large enough to be measured. In addition, the dependence of the Casimir force as a function of $a$ is the expected one from piston geometries.    
Notice that the force is attractive near the the $x=L$ plate and repulsive near the $x=0$ plate.

\subsection*{Small Argument Expansion-Fermion Monopole System}

\begin{figure}[t]
\begin{center}
\includegraphics[scale=.8]{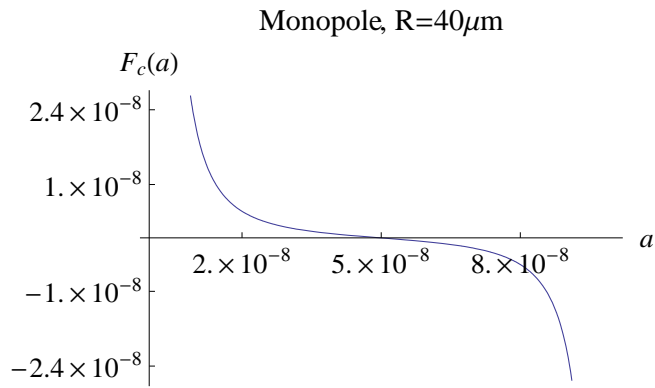}
\end{center}
\caption{The Casimir force on the central plate of a piston, as a function of $a$ for $R=40 \mu$m}
\label{olaskata123}
\end{figure}

In the small argument expansion limit, the behavior of the Casimir force is similar to the large argument approximation, that is, repulsive near the $x=0$ plate and attractive near the $x=L$ plate. This behavior of the Casimir force occurs only for the defect-fermion system and also for the sphere case, as we shall demonstrate in the following sections. For small values of the argument, using relation (\ref{forcesmallq}) and substituting (\ref{encaseorbi}) the Casimir force for the defect-fermion system can be cast as,
\begin{align}\label{forcesmallqmon}
& F_c(s,a)=\frac{5}{2\Gamma(-1/2)}\times
\\ \notag & \sum_{m=0}^{\infty}\frac{(-1)^m\Gamma (m-\frac{3}{2})}{(m-1)!}\zeta_H(-m,\frac{1}{2})\frac{(4\pi)^m}{R^{2m}}\zeta_H\Big{(}2m-3,\frac{1}{2}\Big{)}\Big{(}(L-a)^{2m-4}-a^{2m-4}\Big{)}
\end{align} 
The expression (\ref{forcesmallqmon}) contain terms that converge very fast. 
\begin{center}
\begin{tabular}{|c|c|}
  \hline
  $m$ & $F_c(a)$ \\
  \hline
  0 & 0 \\
  \hline
  1 & -2.10${\,}10^{-8}$ \\
  \hline
  2 & 0 \\
  \hline
  3 & 1.58${\,}10^{-20}$ \\
  \hline
  4 & 0 \\
   \hline
  5 & -1.005${\,}10^{-30}$ \\
   \hline
  6 & 0 \\
   \hline
  7 & 2.69${\,}10^{-40}$ \\
   \hline
  8 & 0 \\
   \hline
  9 & -1.60${\,}10^{-49}$ \\
   \hline
  10 & 0 \\
  \hline
  11 & 1.7${\,}10^{-58}$ \\
   \hline
  12 & 0 \\
\hline
\end{tabular}
\\ \medskip{ \bfseries{Table 1 } }
\end{center}
The contribution of the $m=0$ term is zero and the dominating term is for $m=1$. The series is fast converging and we can see this by looking at Table 1, where we have calculated the first 12 terms of the summation. Note that the series is converging for all the allowed values of $a$, as can easily be checked. As in the large argument approximation case, we take $L=0.1\mathrm{\mu}$m. The constraints from Newton law experiments have to be taken into account, so that $R<0.05$mm. In the following sections, the values of compactification radius will be assumed to be such, so that this constraint is satisfied. Using the same values for $S$ as in the previous case and for $R=40\mu $m, in Fig. (\ref{olaskata123})  we plot the Casimir force on the central plate of the piston, as a function of $a$. Note that the force is attractive near the $x=L$ plate and repulsive near the $x=0$ plate. 
As in the large argument approximation, the Casimir force drastically changes as a function of the compactification radius.

\section{The Fermionic Casimir Force for Various Compact Extra Dimensional Spaces}

In order to have a clear picture of the results of the previous section, we have to compare these to other results coming from various extra dimensional spaces. Hence, in this section we shall investigate the fermionic Casimir energy for the piston geometry for known and well studied extra dimensional spaces. Particularly, we shall study the torus, sphere and circle extra dimensional spaces. Our main objective is certainly to see the behavior of the Casimir force as a function of the number of the extra dimensions, the boundary conditions and also as a function of the compactification scale, and compare qualitatively the results to those of the previous section.

\subsection{The Fermionic Casimir Force Corresponding to a Two Torus Extra Dimensional Space}

\subsection{Stieffel Whitney Class and boundary conditions}

We start by studying the Casimir force of a Dirac fermion in the piston geometry, when the extra dimensional space is a torus $T^2$. The total spacetime is of the form $\mathcal{M}^4\times T^2$. Non-trivial topology of spacetime, has a direct impact on the field configurations that are allowed on the spacetime \cite{isham}. We shall exploit this fact and also the fact that the first
Stieffel class for the spacetime $\mathcal{M}^4\times T^2$, is not trivial. The spacetime is homeomorphic to $S^1\times S^1\times \mathcal{M}^4$. The topological
properties of $S^1\times S^1\times \mathcal{M}^4$, are classified by the first
Stieffel class $H^{1}(S^1\times S^1\times \mathcal{M}^4,Z_{\widetilde{2}})$ which
is isomorphic to the singular (simplicial) cohomology group
${H}_{1}(S^1\times S^1\times \mathcal{M}^4{,Z}_{2})$, owing to the triviality
of the ${Z}_{\widetilde{2}}$ sheaf. The Stieffel class
$H^{1}(S^1\times S^1\times \mathcal{M}^4,{Z}_{\widetilde{2}})={Z}_{2}\times Z_2$,
describes the twisting of a fibre bundle. More accurately, it describes and
classifies the orientability of a bundle globally.
\begin{figure}[h]
\begin{minipage}{15.5pc}
\includegraphics[width=17pc]{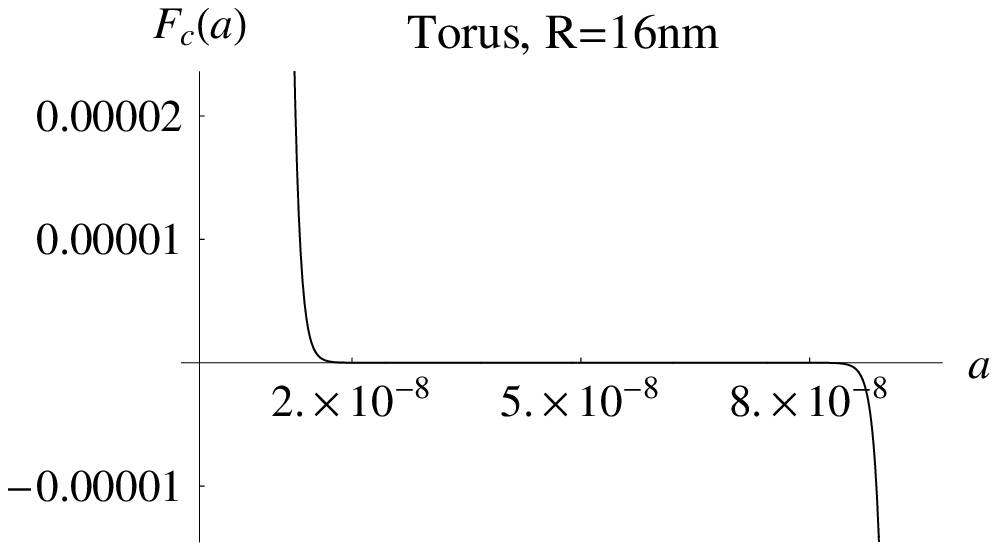}
\end{minipage}
\begin{minipage}{28pc}\hspace{2pc}%
\includegraphics[width=17pc]{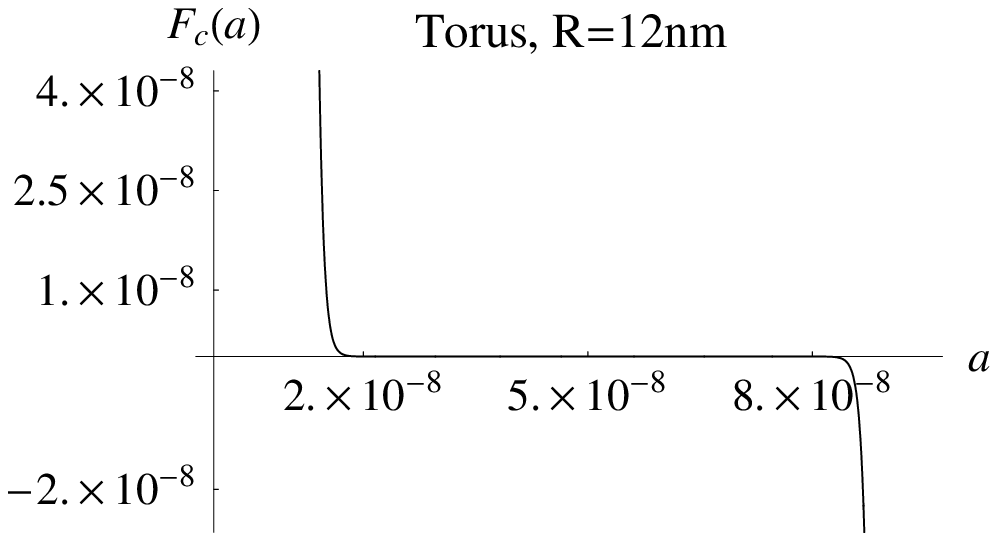}
\end{minipage}
\begin{minipage}[b]{14pc}
\caption{\label{dyoljkljkllkjk}The Casimir force on the central plate of a piston, as a function of $a$ for $R=16$nm and $R=12$nm.}
\end{minipage}
\end{figure}
In the $S^1\times S^1\times \mathcal{M}^4$ case, the
classification group is ${Z}_{2}\times Z_2$ and, we have four locally
equivalent bundles, but different globally. The mathematical
problem at hand, reduces to simply finding the sections that correspond to these four
bundles, classified completely by $Z_{2}\times Z_2$ \cite{isham}. To put it in a different context, these fields satisfy four different boundary conditions in the spacetime $S^1\times S^1\times \mathcal{M}^4$. Usually, due to the Grassman nature of the fermionic fields (and also from finite temperature field theory considerations), one imposes anti-periodic boundary conditions to fermions, disregarding all
other configurations that may arise from non trivial topology. But due to the non-triviality of the Stieffel class, we shall take fermions to obey periodic boundary conditions too. Particularly, we shall focus on the case that the fermions obey only periodic conditions (and disregard combinations such as periodic-anti-periodic etc.). The Kaluza-Klein modes of the fermionic field for periodic boundary conditions in the extra dimensional torus are:  
\begin{equation}\label{encaseorbi1tor}
E_{KK}=\frac{4\pi^2 m^2}{R^2}+\frac{4\pi^2 k^2}{R^2}
\end{equation}
with $m,k=0,\pm 1,...$. Adopting the large argument approximation and the conventions of the previous section, the Casimir force is equal to:
\begin{align}\label{Casflargemonopolegeneral2}
&\mathcal{F}_c\simeq \sum_{n=1}^{\infty}\sum_{m,k=-\infty}^{\infty}\frac{2\pi^{\frac{s-D+1}{2}}\Big{(}\frac{4\pi^2 m^2}{R^2}+\frac{4\pi^2 k^2}{R^2}\Big{)}^{\frac{-4s+2D-2+1}{8}}a^{\frac{-4s-2D-3}{8}}}{\sqrt{n}}n^{s-\frac{D-2}{2}}
 \\  \notag & \times \Big{[}\Big{(}\frac{-4s-2D-3}{8}\Big{)}\frac{1}{a}\Big{(}2^{s-\frac{D-2}{2}+\frac{1}{2}}\exp\Big{(}-
4n\sqrt{\frac{4\pi^2 m^2}{R^2}+\frac{4\pi^2 k^2}{R^2}}a\Big{)}
\\  \notag & -\exp\Big{(}-
2n\sqrt{\frac{4\pi^2 m^2}{R^2}+\frac{4\pi^2 k^2}{R^2}}a\Big{)}
\\  \notag & -\frac{1}{L-a}\Big{(}\frac{-4s-2D-3}{8}\Big{)}-\Big{(}2^{s-\frac{D-2}{2}+\frac{1}{2}}\exp\Big{(}-
4n\sqrt{\frac{4\pi^2 m^2}{R^2}+\frac{4\pi^2 k^2}{R^2}}(L-a)\Big{)}
\\  \notag &-\exp\Big{(}-
2n\sqrt{\frac{4\pi^2 m^2}{R^2}+\frac{4\pi^2 k^2}{R^2}}(L-a)\Big{)}\Big{)}
 \\& \notag-\Big{(}2^{s-\frac{D-2}{2}+\frac{1}{2}}4n\sqrt{\frac{4\pi^2 m^2}{R^2}+\frac{4\pi^2 k^2}{R^2}}\exp\Big{(}-
4n\sqrt{\frac{4\pi^2 m^2}{R^2}+\frac{4\pi^2 k^2}{R^2}}a\Big{)}
\\  \notag &-2n\sqrt{\frac{4\pi^2 m^2}{R^2}+\frac{4\pi^2 k^2}{R^2}}\exp\Big{(}-
2\sqrt{\frac{4\pi^2 m^2}{R^2}+\frac{4\pi^2 k^2}{R^2}}\Big{)}
\\  \notag & -\Big{(}2^{s-\frac{D-2}{2}+\frac{1}{2}}4n\sqrt{\frac{4\pi^2 m^2}{R^2}+\frac{4\pi^2 k^2}{R^2}}\exp\Big{(}-
4n\sqrt{\frac{4\pi^2 m^2}{R^2}+\frac{4\pi^2 k^2}{R^2}}(L-a)\Big{)}
\\  \notag & -2n\sqrt{\frac{4\pi^2 m^2}{R^2}+\frac{4\pi^2 k^2}{R^2}}\exp\Big{(}-
2n\sqrt{\frac{4\pi^2 m^2}{R^2}+\frac{4\pi^2 k^2}{R^2}}(L-a)\Big{)}\Big{)}\Big{]}
\end{align}
In this case, the large argument approximation is valid when the compactification radius of the extra dimensional space is $R \gtrsim 16 $nm, where it is assumed that $L=0.1\mathrm{\mu}$m. Adopting the same values for $S$ and $a$ as in the previous section, and keeping the lowest terms in the summations, we plot in Fig. (\ref{dyoljkljkllkjk}), the dependence of the Casimir force on the central plate, as a function of $a$, for $R=16$nm and $R=12$nm. As in the defect-fermion system a slight change of the compactification radius $R$ results to a huge change of the Casimir force. In addition, the force behaves as the one in the defect-fermion case, that is, attractive near $x=L$ and repulsive near $x=0$. A notable difference from the defect-fermion system, occurs in the small argument approximation as we now demonstrate.

\subsection*{Small Argument Expansion-Double Torus Case}

\begin{figure}[t]
\begin{center}
\includegraphics[scale=.8]{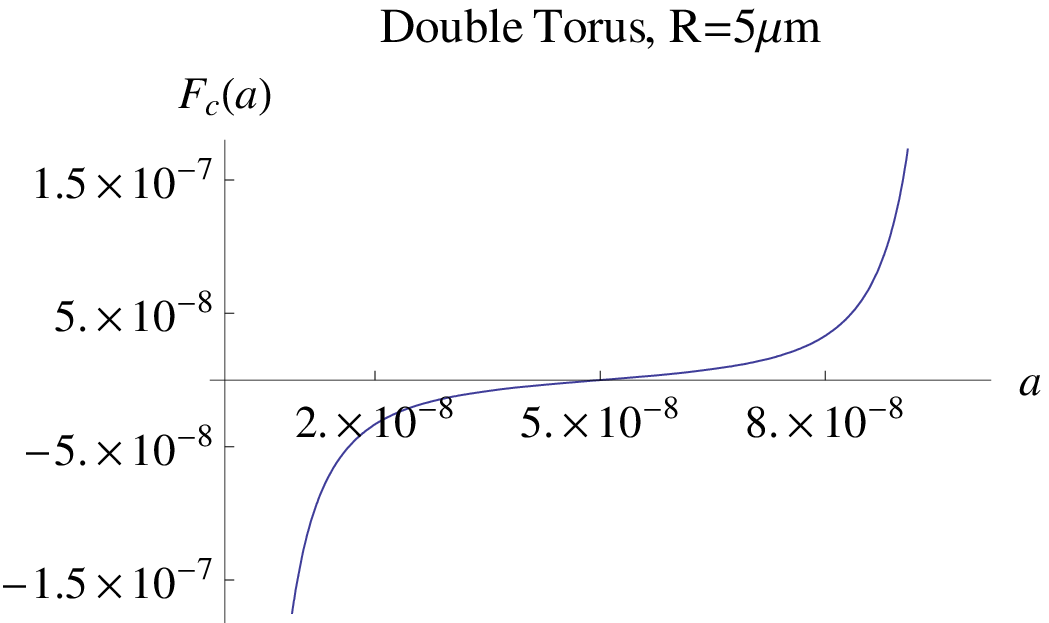}
\end{center}
\caption{The Casimir force on the central plate of a piston, as a function of $a$ for $R=5.5\mathrm{\mu}$m}
\label{olaskatnfnggfgnfa}
\end{figure}

In the small argument expansion limit, using relation (\ref{forcesmallq}) and substituting (\ref{encaseorbi1tor}) the Casimir force for the torus extra dimensional space can be cast in terms of Hurwitz zeta functions and in terms of Dirichlet Beta functions (see in reference \cite{casimirbooks}, Elizalde page 117, relation 5.46),
\begin{align}\label{forcesmallqtordt}
& F_c(s,a)=\frac{5}{2\Gamma(-1/2)}\times
\\ \notag & \sum_{m=0}^{\infty}\frac{(-1)^m\Gamma (m-\frac{3}{2})}{(m-1)!}4\zeta(-m)\beta (-m)\frac{(4\pi)^m}{R^{2m}}\zeta_H\Big{(}2m-3,\frac{1}{2}\Big{)}\Big{(}(L-a)^{2m-4}-a^{2m-4}\Big{)}
\end{align}
with $\beta(x)$, the Dirichlet Beta function, which is equal to \cite{casimirbooks},
\begin{equation}\label{betafiunf}
\beta(x)=\frac{1}{4^x}\Big{(}\zeta_H(x,\frac{1}{4})+\zeta_H(x,\frac{3}{4})\Big{)}
\end{equation} 
The expression (\ref{forcesmallqtordt}) is fast converging, for all the allowed $a$ values, as can be seen in Table 2. The contribution of the $m=0$ term is zero and hence the dominating contribution comes from the $m=1$ term. 
\begin{center}
\begin{tabular}{|c|c|}
  \hline
  $m$ & $F_c(a)$ \\
  \hline
  0 & 0 \\
  \hline
  1 & -1.40${\,}10^{-8}$ \\
  \hline
  2 & 0 \\
  \hline
  3 & -4.23${\,}10^{-21}$ \\
  \hline
  4 & 0 \\
   \hline
  5 & -5.10${\,}10^{-31}$ \\
   \hline
  6 & 0 \\
   \hline
  7 & -5.40${\,}10^{-40}$ \\
   \hline
  8 & 0 \\
   \hline
  9 & -8.34${\,}10^{-47}$ \\
   \hline
  10 & 0 \\
  \hline
  11 & 2.93${\,}10^{-51}$ \\
   \hline
  12 & 0 \\
\hline
\end{tabular}
\\ \medskip{ \bfseries{Table 2 } }
\end{center}
As in the previous cases, we take $L=0.1\mathrm{\mu}$m and $R=40\mu $m. In Fig. (\ref{olaskatnfnggfgnfa})  we plot the Casimir force on the central plate of the piston, as a function of $a$. 
As can be seen from Fig. (\ref{olaskatnfnggfgnfa}), in the small argument approximation, the Casimir force behaves differently, in comparison to the monopole-fermion system, with the force being attractive near the $x=0$ plate and repulsive near the $x=L$ plate.

\subsection{The Fermionic Casimir Force Corresponding to a Sphere Extra Dimensional Space}

\begin{figure}[h]
\begin{minipage}{15.5pc}
\includegraphics[width=17pc]{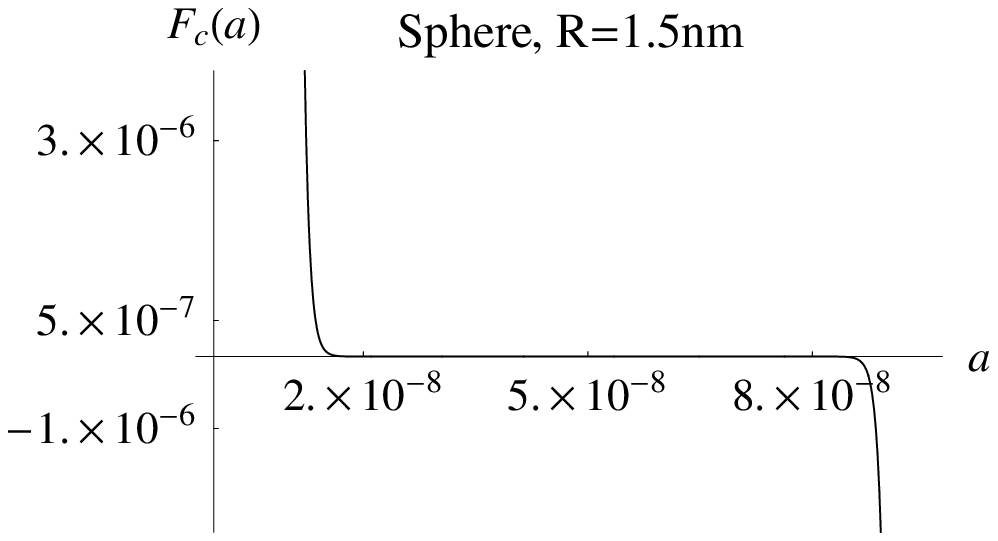}
\end{minipage}
\begin{minipage}{28pc}\hspace{2pc}%
\includegraphics[width=17pc]{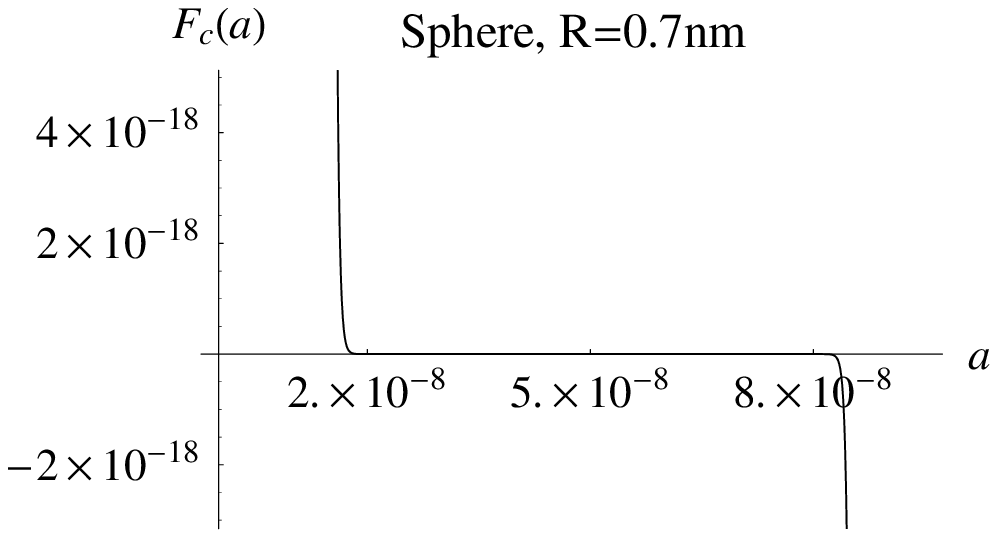}
\end{minipage}
\begin{minipage}[b]{14pc}
\caption{\label{dyo}The Casimir force on the central plate of a piston, as a function of $a$ for $R=1.5$nm and $R=0.7$nm.}
\end{minipage}
\end{figure}

In this subsection the focus is on extra dimensional spaces with sphere topology. The total space is of the form $\mathcal{M}\times S^N$, with $S^N$ denoting the $N$-dimensional sphere. The square of the eigenvalues of the Dirac operator for the $N$-dimensional sphere are of the form \cite{anaforakehagia},
\begin{equation}\label{encaseorbi1}
E_{KK}=\Big{(}m+\frac{N}{2}\Big{)}^2
\end{equation}
with $m=0,1,...$. Following the line of argument of the previous sections, the Casimir force in the large argument approximation reads,
\begin{align}\label{Casflargemonopolegeneral4}
&\mathcal{F}_c\simeq \sum_{n=1}^{\infty}\sum_{m=0}^{\infty}\frac{2\pi^{\frac{s-D+1}{2}}\Big{(}\frac{m+\frac{N}{2}}{R^2}\Big{)}^{\frac{-4s+2D-2+1}{8}}a^{\frac{-4s-2D-3}{8}}}{\sqrt{n}}n^{s-\frac{D-2}{2}}
 \\  \notag & \times \Big{[}\Big{(}\frac{-4s-2D-3}{8}\Big{)}\frac{1}{a}\Big{(}2^{s-\frac{D-2}{2}+\frac{1}{2}}\exp\Big{(}-
\frac{4n\Big{(}m+\frac{N}{2}\Big{)}}{R}a\Big{)}
\\  \notag & -\exp\Big{(}-
\frac{2n\Big{(}m+\frac{N}{2}\Big{)}}{R}a\Big{)}
\\  \notag & -\frac{1}{L-a}\Big{(}\frac{-4s-2D-3}{8}\Big{)}-\Big{(}2^{s-\frac{D-2}{2}+\frac{1}{2}}\exp\Big{(}-
\frac{4n\Big{(}m+\frac{N}{2}\Big{)}}{R}(L-a)\Big{)}
\\  \notag &-\exp\Big{(}-
\frac{2n\Big{(}m+\frac{N}{2}\Big{)}}{R}(L-a)\Big{)}\Big{)}
 -\Big{(}2^{s-\frac{D-2}{2}+\frac{1}{2}}\frac{4n\Big{(}m+\frac{N}{2}\Big{)}}{R}\exp\Big{(}-
\frac{4n\Big{(}m+\frac{N}{2}\Big{)}}{R}a\Big{)}
\\  \notag &-\frac{2n\Big{(}m+\frac{N}{2}\Big{)}}{R}\exp\Big{(}-
2\frac{\Big{(}m+\frac{N}{2}\Big{)}}{R}\Big{)}
\\  \notag & -\Big{(}2^{s-\frac{D-2}{2}+\frac{1}{2}}\frac{4n\Big{(}m+\frac{N}{2}\Big{)}}{R}\exp\Big{(}-
\frac{4n\Big{(}m+\frac{N}{2}\Big{)}}{R}(L-a)\Big{)}
\\  \notag & -\frac{2n\Big{(}m+\frac{N}{2}\Big{)}}{R}\exp\Big{(}-
\frac{2n\Big{(}m+\frac{N}{2}\Big{)}}{R}(L-a)\Big{)}\Big{)}\Big{]}
\end{align}
Using the same values for $L$, and $S$ as in the previous, this approximation is valid when the compactification radius is $R\lesssim 1.5$nm. In Fig. (\ref{dyo}), we plot the Casimir force as a function of $a$, for two values of the compactification radius, namely for $R=0.7$nm and $R=1.5$nm, and also for $N=2$. 
We can observe that the Casimir force in the case which the extra space is a 2-sphere, behaves as the torus and defect-fermion space behave. Hence, we can see that a slight change in the compactification radius, can cause huge changes in the Casimir force. 
\subsection*{Small Argument Expansion--2-Sphere Case}

\begin{figure}[t]
\begin{center}
\includegraphics[scale=.8]{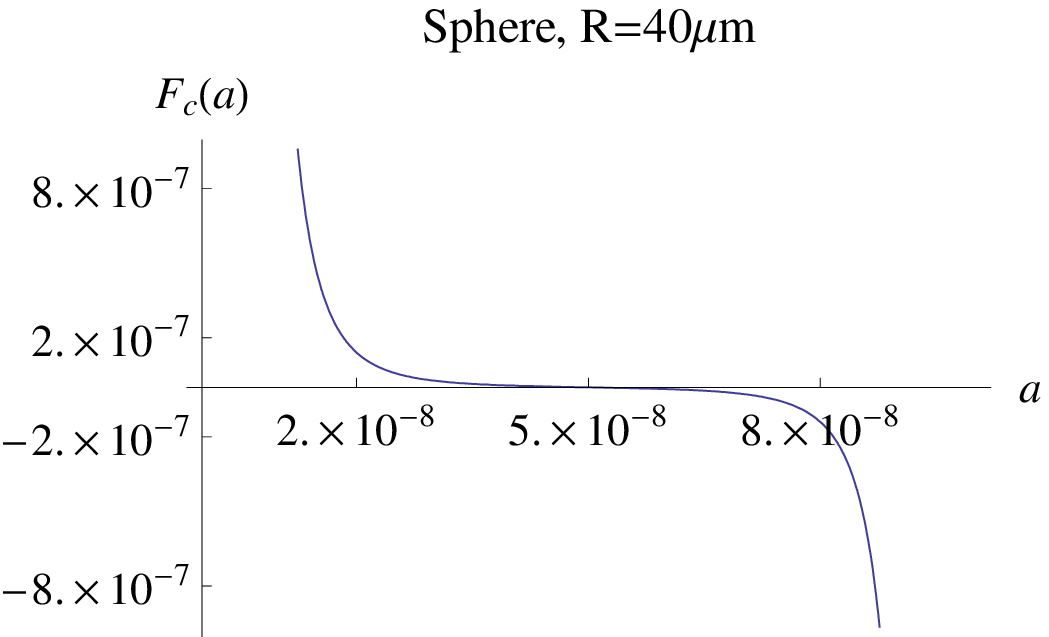}
\end{center}
\caption{The Casimir force on the central plate of a piston, as a function of $a$ for $R=3\mathrm{\mu}$m}
\label{olaskathgfhfghfa}
\end{figure}

Accordingly, in the small argument approximation, the Casimir force for the N-sphere case takes the following form,
\begin{align}\label{forcesmallqsphere}
& F_c(s,a)=\frac{5}{2\Gamma(-1/2)}\times
\\ \notag & \sum_{m=0}^{\infty}\frac{(-1)^m\Gamma (m-\frac{3}{2})}{(m-1)!}4\zeta_H(-2m,\frac{N}{2})\frac{1}{R^{2m}}\zeta_H\Big{(}2m-3,\frac{1}{2}\Big{)}\Big{(}(L-a)^{2m-4}-a^{2m-4}\Big{)}
\end{align}
The expression (\ref{forcesmallqsphere}) is particularly simple, since for the 2-sphere case, we have $\zeta_H(-2m,1)=0$, $\forall$ $m>0$. Hence, the only non zero contribution comes from the $m=0$ term, which is $\zeta_H(0,1)=-0.5$. For $L=0.1\mathrm{\mu}$m and $R=40\mu $m, in Fig. (\ref{olaskathgfhfghfa})  we plot the Casimir force on the central plate of the piston, as a function of $a$. 
As can be seen from Fig. (\ref{olaskathgfhfghfa}), in the small argument approximation, the Casimir force behaves in exactly the same way as the force in the case of the monopole-fermion system, with the force being repulsive near the $x=0$ plate and attractive near the $x=L$ plate.

\subsection{The Fermionic Casimir Force Corresponding to a Circle Extra Dimensional Space}

As we demonstrated in the previous sections, the fermionic Casimir force of the defect-fermion system and for a sphere extra space, confined in a piston geometry, behaves very differently in comparison to the double torus extra dimensional space. The comparison was made with spaces having different topology, in six dimensions. This different behavior may be owing to the space dimensionality, and hence it worths to briefly present a five dimensional extra space in order to see what happens in this case. We shall assume that the extra dimensional space is a circle $S^1$ of radius $R$. The total spacetime is of the form $\mathcal{M}^4\times S^1$.

\begin{figure}[h]
\begin{minipage}{15.5pc}
\includegraphics[width=17pc]{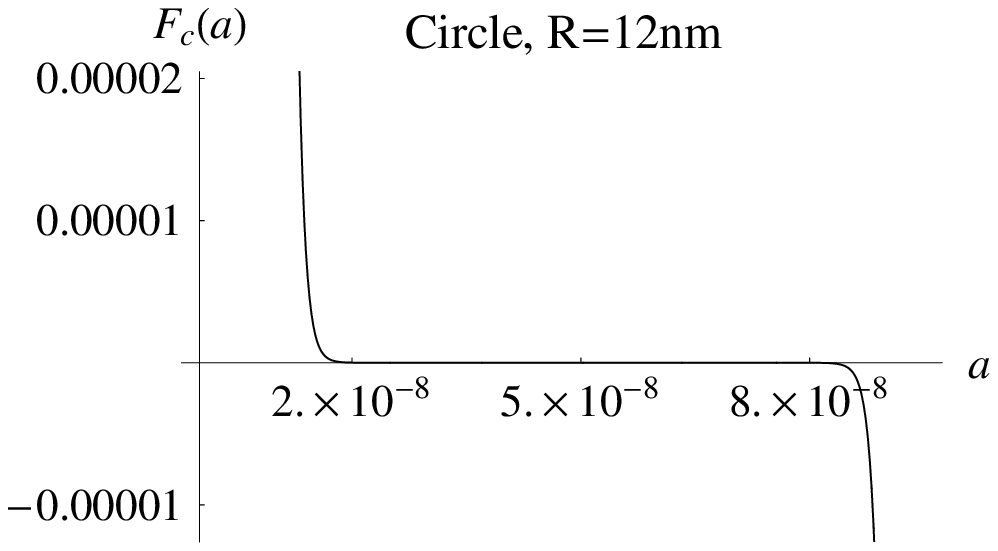}
\end{minipage}
\begin{minipage}{28pc}\hspace{2pc}%
\includegraphics[width=17pc]{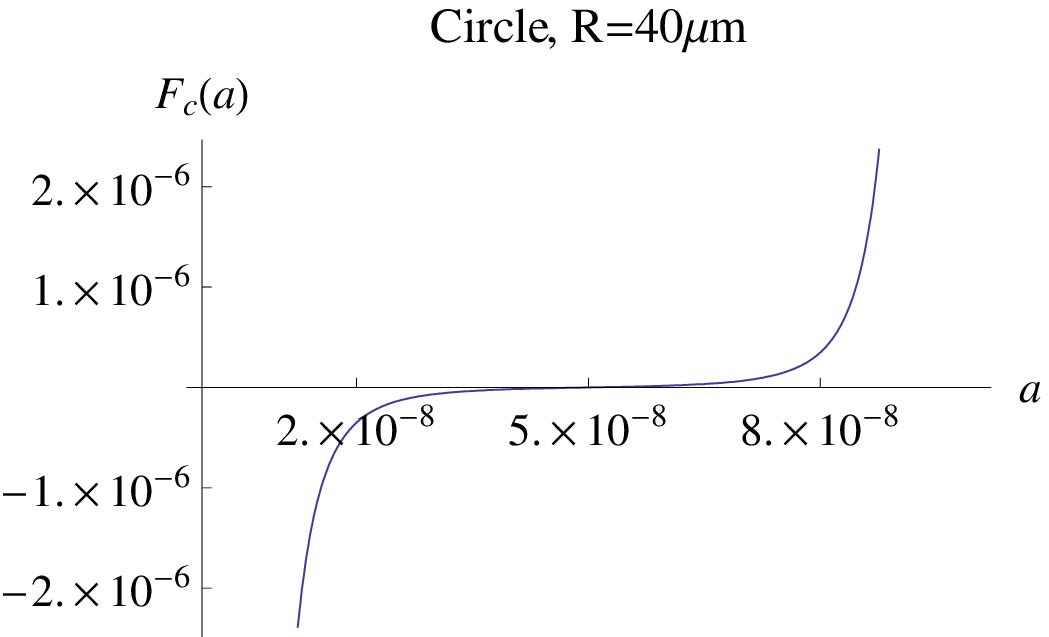}
\end{minipage}
\begin{minipage}[b]{14pc}
\caption{\label{dyjljlljo} The Casimir force as a function of $a$. On the left $R=12$nm and on the right $R=40\mathcal{\mu}$m.}
\end{minipage}
\end{figure}

As in the torus case, the topological
properties of $ S^1\times \mathcal{M}^4$, are classified by the first
Stieffel class $H^{1}(S^1\times \mathcal{M}^4,Z_{\widetilde{2}})$ which
in this case is equal to $Z_2$. This implies that the total number of fermionic sections allowed for this spacetime is 2. Actually, this means that the fermionic sections can satisfy periodic and anti-periodic boundary conditions. We choose the case for which the fermions are periodic (the other case yields similar results). The Kaluza-Klein modes of the Dirac operator for the circle, $ S^1$ (with periodic boundary conditions for the fermions), are 
\begin{equation}\label{encaseorbi1treyryr}
E_{KK}=\frac{4\pi^2 m^2}{R^2}
\end{equation}
with $m=0,\pm 1,\pm 2,...$. Using the same values for $S$ and $L$, the large argument approximation is valid when $R\lesssim 12$nm, while the small argument approximation is valid when $R\sim 40\mathcal{\mu}$m. In the left figure of Fig. (\ref{dyjljlljo}),
we plot the Casimir force in the large argument approximation, where we observe that the force behaves similarly to the other extra dimensional spaces. In the small argument case, the Casimir force is equal to:
\begin{align}\label{forcesmallqtor}
& F_c(s,a)=\frac{5}{2\Gamma(-1/2)}\times
\\ \notag & \sum_{m=0}^{\infty}\frac{(-1)^m\Gamma (m-\frac{3}{2})}{(m-1)!}\Big{(}1+\frac{\zeta (-2m)}{2}\Big{)}\frac{(2\pi)^{2m}}{R^{2m}}\zeta_H\Big{(}2m-3,\frac{1}{2}\Big{)}\Big{(}(L-a)^{2m-4}-a^{2m-4}\Big{)}
\end{align} 
The expression (\ref{forcesmallqtor}) is fast converging, for all the allowed $a$ values, as can be seen in Table 3. The dominating contribution comes from the $m=0$ term. In the right figure of Fig. (\ref{dyjljlljo}), we have plotted the dependence of the small argument approximation Casimir force of the circle, as a function of $a$. Note that in this case, as in the case of the torus, the Casimir force is attractive at the $x=0$ plate and repulsive at the $x=L$ case. 

\begin{center}
\begin{tabular}{|c|c|}
  \hline
  $m$ & $F_c(a)$ \\
  \hline
  0 & -5.61${\,}10^{-6}$ \\
  \hline
  1 & 5.04${\,}10^{-12}$ \\
  \hline
  2 & 0 \\
  \hline
  3 & -2.03${\,}10^{-28}$ \\
  \hline
  4 & 2.61${\,}10^{-33}$ \\
   \hline
  5 & -3.93${\,}10^{-38}$ \\
   \hline
  6 & 6.51${\,}10^{-43}$ \\
   \hline
  7 & -1.15${\,}10^{-47}$ \\
   \hline
  8 & 2.15${\,}10^{-52}$ \\
   \hline
  9 & -4.15${\,}10^{-57}$ \\
   \hline
  10 & 8.62${\,}10^{-62}$ \\
  \hline
  11 & -1.69${\,}10^{-66}$ \\
   \hline
  12 & 1.15${\,}10^{-72}$ \\
\hline
\end{tabular}
\\ \medskip{ \bfseries{Table 3} }
\end{center}

\section{Discussion and Concluding Remarks}

In this paper we calculated the Casimir force for fermionic quantum fields confined in a piston that consists of three parallel plates. We assumed that the field satisfies bag boundary conditions on the plates, and also that the field is massless. In addition, we also assumed that the spacetime has compact extra dimensions and we studied how these affect the fermionic Casimir force on the central plate. Particularly, we examined four different extra spaces, namely, a sphere, a torus, a circle and a torus with the existence of a non-trivial defect in the extra space coupled to the fermion. 

\noindent The Casimir force was calculated for these extra spaces in two limiting cases, that is, in the large argument limit and in the small argument limit, with the argument being dependent on the fraction $a/R$. Quantitatively, if the distances of the plate are of the order of $0.1\mathcal{\mu}$m, the large argument limit is achieved when the compactification radius of the extra space is of the order $10$nm, while the small argument limit is achieved when the compactification radius is lower than the Newton law experiments limit  $R<50 \mu$m. As we explicitly demonstrated, the case in which the extra space contains a defect coupled to the fermion and also when the extra space is a sphere, in the small argument limit, the Casimir force has opposite sign compared to the Casimir force of the toroidal topologies.

\noindent Apart from this apparent difference between the aforementioned spaces, the Casimir force in the large argument limit behaves in the same way for all the spacetimes. A general feature of the Casimir force for all cases, is that a small change in the compactification radius amounts in huge changes in the Casimir force. We assumed that the compactification radius is smaller than $R=0.05$mm. This constraint is imposed by Newton law experiments, in which deviations from the standard Newton law due to extra dimensions are measured.

In the present study, we tried to have a hint on how the fermionic Casimir energy behaves, for a piston geometry with three parallel plates. We used realistic length scales in our formulas but the results are by far incomplete. This is owing to the fact that a realistic calculation should take into account the finite temperature corrections, roughness corrections and also other issues that could critically change the result \cite{emig}. However, the outcomes of this paper show us some very interesting features of the Casimir force in the presence of extra dimensions. Specifically, the measurement of the Casimir force can show us which extra space structure causes this force. This is in contrast to Newton law experiments, in which a phenomenon called shadowing can make the identification of the extra space, almost impossible \cite{oikonomounewton}. Thus, in conjunction with the fact that the Casimir effect experiments are less expensive than particle accelerators experiments, these can serve as a tool for revealing the micro-structure of spacetime and also for unveiling the quantum structure (if any) of spacetime.

\noindent Furthermore, the Casimir effect is important in the construction and operation of nano-scale and micro-scale devices. Therefore, a better understanding of the Casimir force dependence on shape, material and geometry is of crucial importance \cite{emig}. The Casimir force is one of the principal causes of malfunctions in these devices because of the stiction of nearby elements. The van der Waals forces have an important property, the fact that these are non-additive, which gives a many body interpretation of the forces.  Hence, a better theoretical understanding of the Casimir-van der Waals forces is of particular importance. In addition, owing to the fact that the Casimir forces increase strongly when the distance between elements of the nano-devices decrease, these forces play a crucial role in devices that contain nano-scale or (micro-scale) moving elements, the so called nano-scale (micro-scale) actuators. Hence, a minimum separation between the elements must be determined in order to avoid stiction. In addition, in periodically structured parallel surfaces, the Casimir force between the surfaces can be controlled. Notice that, the piston geometry is analogous to these periodical parallel surfaces construction, hence the present study might offer valuable information for these devices.

\end{document}